\documentclass{iopart}
\usepackage{graphicx}

\newcommand \beq{\begin{eqnarray}}
\newcommand \eeq{\end{eqnarray}}
\newcommand{\mnote}[1]{\marginpar{\tiny {}}}   

\def \ee   {\mbox{e$^+$e$^-$}}

\begin{document}

\title{Particle Ratios, Equilibration, and the QCD Phase Boundary}
\author{P. Braun-Munzinger}
\address{Gesellschaft f\"ur Schwerionenforschung, D-64291 Darmstadt, Germany}

\author{J. Stachel}
\address{Physikalisches Institut der Universit\"at Heidelberg, D 69120
Heidelberg, Germany}

\begin{abstract}

    We discuss the status of  thermal model descriptions of
    particle ratios in central  nucleus-nucleus collisions at
    ultra-relativistic energy. An alternative to the
    ``Cleymans-Redlich'' interpretation 
    of the freeze-out  trajectory is given in terms of the total
    baryon density. Emphasis is placed on the relation
    between the chemical equilibration parameters and the QCD phase
    boundary. Furthermore, we trace  the essential difference between
    thermal model analyses of data from collisions between elementary
    particles and from heavy ion collisions as due to a transition
    from local strangeness conservation to percolation of strangeness
    over large volumes, as occurs naturally in a deconfined medium.
    We also discuss predictions of the thermal model for composite
    particle production.

\end{abstract}

The thermal model has been used very successfully to reproduce and
(for RHIC data even) predict particle ratios measured in central
collisions of heavy nuclei at ultra-relativistic energy
\cite{therm1,therm2,therm3,becattini,cleymans,pbm_rhic}. In
particular, the predictions for the RHIC data are very close to the
experimental observations: for a temperature of T=174 MeV and a baryon
chemical potential of $\mu_b=$ 46 MeV all available data measured near
central rapidity are reproduced with a $\chi^2$ of 5.7 for 7 effective
degrees of freedom \cite{pbm_rhic}. We note that these calculations
are performed under the condition of full strangeness equilibration
and global strangeness conservation, i.e. the flow of strangeness into
and out of the central rapidity slice is assumed to cancel.  For an
update on the inclusion of the most recent data see \cite{magestro},
where it is demonstrated that even the $\phi/$h$^-$ ratio is well
described with the same set of parameters.

\begin{figure}[t]
\begin{center}
\includegraphics[width=10cm]{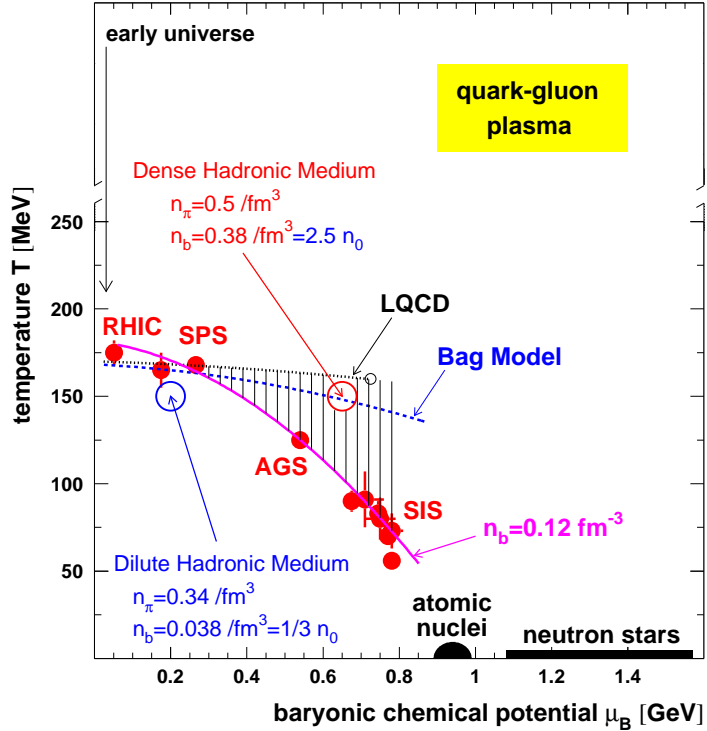}
\end{center}
\caption{The QCD Phase Diagram in the context of the current knowledge about
relativistic nuclear collisions. For details see text.}
\label{fig:phase}
\end{figure}

These analyses yield a parameter pair (T, $\mu_b$) for each energy
where data have been obtained. The current state of affairs is
described in Fig. ~\ref{fig:phase}.  In this plot of temperature vs
chemical potential, the location of the phase boundary between
hadronic matter and the quark-gluon plasma is indicated by the dotted
and dashed lines, taken from the latest calculations within the
framework of lattice QCD \cite{fodor} and from a bag model equation of
state \cite{gerry_70}. The data points in this figure represent
chemical freeze-out points determined by a thermal analysis
\cite{therm1,therm2,therm3,pbm_rhic,pbmqm97} of hadron multiplicities
measured in central Pb+Pb or Au+Au collisions at various beam energies
at the SIS, AGS, SPS and RHIC accelerator facilities. The line
connecting the data points demonstrates, in the framework of the
hadron resonance gas model used in \cite{therm3}, that chemical
freeze-out appears to take place at constant total baryon plus
anti-baryon density n$_b = 0.12/{\rm fm}^3$. The so determined
freeze-out line  crosses the calculated phase boundary
roughly at full SPS energy.

We note that this interpretation of the freeze-out trajectory differs
significantly from that of \cite{cleymans}, where chemical freeze-out
is assumed to take place at constant total energy per particle
$<E>/<N> \approx 1$ GeV. To us the constant total baryon density
scenario is more appealing as it implies a physical picture: chemical
freeze-out takes place at a critical baryon density through
baryon-baryon and baryon meson-interactions. The absolute value of the
critical baryon density  n$_b = 0.12/{\rm fm}^3$ depends of course on
the fact that, in \cite{therm3}, a repulsive hadron-hadron interaction
is implemented via an excluded volume.

The fact that the chemical freeze-out points closely approach the
calculated QCD phase boundary lends strong support to the
interpretation \cite{gerry_70,stock,heinz,jsinpc,pbmpanic} that the matter 
produced in nuclear collisions at SPS and RHIC energies was first
thermalized in the deconfined  quark-gluon plasma  phase and
subsequently expanded through the phase boundary into a thermal gas of
(weakly interacting) hadrons. 

Furthermore, as is illustrated in Fig. ~\ref{fig:phase}, the hadronic
matter underneath the phase boundary for values of the baryon chemical
potential $\mu_b < 300 $ MeV, roughly corresponding to full SPS energy
and higher, is always dilute: the total baryon density never exceeds
that of normal nuclear matter. We, consequently, do not expect a
strong influence on the chemical freeze-out scenario due to changed
hadron masses in the medium, if SPS and RHIC energies are
considered. Furthermore, even if one includes very strong medium effects
such as in the chiral model of \cite{zschiesche}, the chemical
freeze-out still occurs close to the phase boundary. In the analysis
of \cite{zschiesche} the phase boundary is at T$_c$ = 154 MeV, while
the chiral model analysis of the data yields T=144 MeV.

It is only at lower beam energies, between 5 and 50
GeV/nucleon, that one can identify between chemical freeze-out and the
phase boundary a phase of dense hadronic matter, shown by the hatched
area in Fig. ~\ref{fig:phase}. It is near these beam energies,
i.e. close to 30 GeV/nucleon, where also the relative strangeness
content of the hadronic medium is maximum \cite{strange_max}.

This interpretation calls into question the widely accepted belief
that the enhanced yield of low mass dileptons observed at full SPS
energy by the CERES collaboration \cite{ceres1,ceres2} is mainly due
to interactions of the $\rho$-meson with baryons \cite{rapp}. 

At 40 GeV/nucleon beam energy, analysis \cite{schmitz} of the (still
preliminary) hadron yields from the NA45, NA49, and NA57
collaborations leads to freeze-out parameters of T$\approx 140$ MeV
and $\mu_b \approx 400$ MeV. This $\mu_b$ value is very close to the
value of about 420 MeV obtained from the phenomenological
parameterization provided by equation 1 in \cite{strange_max} at 40
GeV/nucleon. At this energy, the CERES collaboration has recently
provided first, preliminary results on low-mass di-electrons
\cite{damja,fili}. Again, an enhancement relative to the yield
expected from standard hadronic sources is observed. In fact, the
observed enhancement seems to increase from 2.7 $\pm 0.65$ at 160
GeV/nucleon to 4.5 $\pm 1.2$ at 40  GeV/nucleon. Whether this is due
to the increase, at 40 compared to 160 GeV/nucleon, of the baryon
density between chemical freeze-out and the phase boundary is an
interesting speculation. However, if Fig. ~\ref{fig:phase} represents
reality, this increase is only slight, and the real effect should
develop at around 20 GeV/nucleon. Whether the few days of running
currently planned at the CERN SPS in 2002 at 30 and 20 GeV/nucleon will lead to
further insights into this intriguing question is unclear at present.

We note that the planned new GSI accelerator facility \cite{gsi} is
tailored to explore in detail the interesting region between 5 and 30
GeV/nucleon, where a dense hadronic medium is expected between the QCD
phase boundary and chemical freeze-out.

The above discussed connection between the QCD phase boundary and the
observed chemical freeze-out points is sometimes called into question
\cite{specht,nagle} since hadron production in \ee and pp or 
p$\bar{\rm{p}}$ can also be described in thermal model \cite{becattini1}
yielding a (apparently universal) temperature T$_e \approx 170$
MeV. While the fact that T$_e$ is close to T values determined for
heavy ion collisions at top SPS and RHIC energies (corresponding to
relatively small values of $\mu_b$) might indeed reflect the
fundamental hadronization scale of QCD, we note that there is an
essential difference between thermal descriptions of central heavy ion
collisions and elementary particle reactions. As discussed, particle
densities and their ratios can be, for heavy ion collisions at full
AGS energy and higher, well described in the grand-canonical ensemble
using the particle density $\rm{n_{gc}}$ . 
In contrast, for a description of particle production in elementary
particle collisions, local quantum number conservation needs to be
taken into account explicitly, leading, for the production of
particles with strangeness $n$ to a canonical suppression factor \beq
F_c = \frac{I_n( x_1)}{I_0( x_1)}, \eeq with which the
grand-canonical densities have to be multiplied to obtain the results
in the canonical ensemble. Here, $I_n$ denotes a modified Bessel
function of degree $n$ and the parameter $x_1 = 2 \rm{V} \rm{n_{gc}}$
depends explicitly on the volume V over which quantum number
conservation is enforced \cite{redlich}. For $x_1 >> 1$, $\frac{I_n(
x_1)}{I_0( x_1)} \rightarrow 1$, the volume drops out and the
grand-canonical limit applicable to ultra-relativistic nuclear
collisions is recovered. For $x_1 << 1, \frac{I_n( x_1)}{I_0( x_1)}
\rightarrow \frac{(1/2\cdot x_1)^n}{n!}$. This case, appropriate for
the analysis of \ee and pp or p$\bar{\rm{p}}$ collisions, implies that
densities of particles with strangeness $n$ contain a quantum number
dependent volume V. Hence, the K/$\pi$ ratios are proportional to V,
the $\Omega/\pi$ ratios proportional to V$^3$. For the description of
elementary particle collisions V = V$_0 \approx 7$ fm$^3$
\cite{redlich} is required, i.e. strangeness is conserved within
approximately the volume of a nucleon.  Moving to an analysis of heavy
ion collisions one finds \cite{strange_max,redlich} that V = V$_0
\cdot$ N$_{part}$/2 where N$_{part}$ is the number of
participants. Already in central Au-Au collisions at SIS energies this
``correlation volume'' V exceeds 1000 fm$^3$. The required volume
increases further as the beam energy is increased, until the
grand-canonical limit is reached at full SPS energy.

This implies that, in central nucleus-nucleus collisions at
ultra-relativistic energies, strangeness percolates freely over
volumes of thousands of fm$^3$! At top SPS and RHIC energies it is
natural to conclude that the percolation has its origin in the
quark-gluon phase, lending further strong support to the
interpretation above that the ``coincidence'' between experimentally
determined chemical freeze-out points and the calculated phase
boundary implies that a deconfined phase was produced in such
collisions. An interesting question is what leads to the percolation
at SIS energies, where the freeze-out points are far away from where
we think the phase boundary is. We speculate that, in this case, the
percolation takes place in the high density hadronic phase indicated
by the hatched area in Fig. ~\ref{fig:phase}.

An often overlooked aspect of the thermal model is the possibility to
compute also the yields of composite particles. For example, the d/p
and $\bar{\rm{d}}/\bar{\rm{p}}$ ratios measured at SPS and AGS
energies are well reproduced with the same parameters which are used
to describe baryon and meson ratios
\cite{therm1,therm2,therm3}. Furthermore, the AGS E864 collaboration
has recently published \cite{E864} yields for composite particles
(light nuclei) produced in central Au-Au collisions at AGS energy near
mid-rapidity. In this investigation, an exponential decrease of
composite particle yield with mass is observed, implying a penalty
factor of about 48 for each additional nucleon. Extrapolation of the
data to large transverse momentum values reduces this penalty factor
to about 26, principally because of transverse flow. In the thermal
model, this penalty factor can be easily derived. In the relevant
Boltzmann approximation, we obtain

\beq
R_p \approx exp{\frac{m \pm \mu_b}{T}},
\eeq

\noindent where m is the nucleon mass and the negative sign applies
for matter, the positive for anti-matter. Small corrections due to the
spin degeneracy and the A$^{3/2}$ term in front of the exponential in
the Boltzmann formula for particle density are neglected.  Using the
freeze-out parameters T=125 MeV and $\mu_b = $ 540 MeV appropriate for
AGS energy \cite{therm1,therm3} we calculate R$_p \approx 23$, in
close agreement with the data for the production of light nuclei. We
also note that the anti-matter yields measured by the E864
collaboration \cite{E864_1} yield penalty factors of about 2$\cdot
10^5$, again close to the predicted value of 1.3$\cdot 10^5$.

This rather quantitative agreement between measured relative yields  for
composite particles and thermal model predictions  provides some
confidence in the predictions for yields of exotic objects produced in
central nuclear collisions. We briefly comment here on the results of
a 1995 analysis \cite{jphysg}.

In this investigation, the production probabilities for exotic strange
objects and, in particular, for strangelets were computed in the
thermal model. The results are reproduced in Table\ \ref{probab} for
temperatures relevant for beam energies between 10 and 40
GeV/nucleon. We first note that predictions of the thermal model and,
where available, the coalescence model of \cite{dover1} agree (maybe
surprisingly) well. Secondly, inspection of Table\ \ref{probab}
also shows that, in future high statistics experiments which will be
possible at the planned new GSI facility \cite{gsi}, multi-strange
objects such as $_{\Xi^0\Lambda \Lambda}^7$He should be experimentally
accessible with a planned sensitivity of about 10$^{-13}$ per central
collision in a years running, should they exist and be produced with
thermal yields. Investigation of yields of even the lightest
conceivable strangelets will be difficult, though.

\begin{table}
\caption{Produced number of nonstrange and strange clusters and
of strange quark 
matter  per  central Au+Au collisions at AGS energy, calculated in a
thermal model for two different 
temperatures, baryon chemical potential $\mu_b$= 0.54 GeV and
strangeness chemical potential $\mu_s$ such that overall strangeness is
conserved. The coalescence model predictions are from Table 2 of \cite{dover1}.
\label{probab}}
\begin{indented}
\item[]\begin{tabular}{@{}llll}
\br
 & \centre{2}{Thermal Model} & \\
\ns
 & \crule{2} & \\
Particles & $T$=.120 GeV & $T$=.140 GeV & Coalescence Model\\ 
\mr 
d & 15 & 19 & 11.7  \\
t+$^3$He & 1.5 & 3.0 & 0.8  \\
$\alpha$ & 0.02 & 0.067 & 0.018 \\
$H_0$       & 0.09 & 0.15 & 0.07  \\
$_{\Lambda \Lambda}^5$H & 3.5 $\cdot 10^{-5}$ & 2.3 $\cdot 10^{-4}$ &
4$\cdot 10^{-4}$ \\ 
$_{\Lambda \Lambda}^6$He & 7.2 $\cdot 10^{-7}$ & 7.6 $\cdot 10^{-6}$ & 
 1.6$\cdot 10^{-5}$ \\
$_{\Xi^0\Lambda \Lambda}^7$He & 4.0 $\cdot 10^{-10}$ & 9.6 $\cdot
10^{-9}$  & 4 $\cdot 10^{-8}$  \\ \mr
$^{10}_{1}$St$^{-8}$ &  1.6 $\cdot 10^{-14}$ & 7.3 $\cdot 10^{-13}$ & \\
$^{12}_1$St$^{-9}$ &  1.6 $\cdot 10^{-17}$ & 1.7 $\cdot 10^{-15}$  &  \\
$^{14}_1$St$^{-11}$ & 6.2 $\cdot 10^{-21}$ & 1.4 $\cdot 10^{-18}$  & \\
$^{16}_2$St$^{-13}$ & 2.4 $\cdot 10^{-24}$ & 1.2 $\cdot 10^{-21}$   & \\
$^{20}_2$St$^{-16}$ & 9.6 $\cdot 10^{-31}$ & 2.3 $\cdot 10^{-27}$   & \\
\br
\end{tabular}
\end{indented}
\end{table}

In summary, we have discussed the surprising success of the thermal
model to reproduce particle yields in central nucleus-nucleus
collisions at ultra-relativistic energies. These investigations yield
chemical freeze-out parameters which approach the phase boundary for
energies at and above full SPS energy. The implications of this
observation are, as discussed above, that the matter produced in
nuclear collisions at SPS and RHIC energies was most likely first
thermalized in a deconfined quark-gluon plasma phase and subsequently
expanded through the phase boundary into a gas of hadrons. In a
detailed discussion of various areas of the phase diagram we have
identified a zone of dense hadronic matter between the expected phase
boundary and chemical freeze-out which can be studied experimentally
with beams in the energy region 10 $<$ E$_{lab}$/A $<$ 40
GeV. Furthermore, we have traced the essential difference between
thermal model analyses of data from collisions between elementary
particles and from heavy ion collisions as due to a transition from
local strangeness conservation to percolation of strangeness over
large volumes, as occurs naturally in a deconfined medium.  Finally, a
discussion of composite particle production has emphasized the success
of the thermal model also in this area and indicated possibilities to
study experimentally also rather exotic and interesting multi-strange
hadrons.


\section*{References}
\begin{thebibliography}{9}

\bibitem{therm1} Braun-Munzinger P, Stachel J, Wessels J P and Xu N
1995 \PL B {\bf 344} 43

\bibitem{therm2} Braun-Munzinger P, Stachel J, Wessels J P and
Xu N 1996 \PL B {\bf 365} 1
 
\bibitem{therm3} Braun-Munzinger P, Heppe I and Stachel J
1999 \PL B {\bf 465} 15

\bibitem{becattini} Becattini F, Gazdzicki M and Sollfrank J
1998 \EJP C {\bf 5} 143

\bibitem{cleymans} Cleymans J and Redlich K 1999 \PR C {\bf 60} 054908

\bibitem{pbm_rhic} Braun-Munzinger P, Magestro D, Redlich K and
Stachel J 2001 \PL B {\bf 518} 41\\
({\it Preprint} hep-ph/0105229)
 
\bibitem{magestro} Magestro D 2001 {\it these proceedings}\\
({\it Preprint} hep-ph/0112178)

\bibitem{fodor} Fodor Z and Katz S D 2001 {\it Preprint} hep-lat/0106002

\bibitem{gerry_70} Braun-Munzinger P and Stachel J
1996 \NP A {\bf 606} 320 \\ Note that, in Fig. ~\ref{fig:phase}, 
the dashed line is calculated 
with a rescaled bag constant such that T$_c$ = 167 MeV for $\mu_b = 0$. 

\bibitem{pbmqm97} Braun-Munzinger P and Stachel J 1998 \NP A {\bf 638}
3c

\bibitem{stock} Stock R 1999 \PL B {\bf 456} 277

\bibitem{heinz} Heinz U 1999 \NP A {\bf 661} 140c

\bibitem{jsinpc} Stachel J 1999 {\it Proc. INPC, Paris, August 1998}
\NP A {\bf 654} 119c

\bibitem{pbmpanic} Braun-Munzinger P 2000 {\it Proc. PANIC, Uppsala, June
1999} \NP A {\bf 663-664} 183

\bibitem{zschiesche} Zschiesche D {\it et al} 2001 \NP A {\bf
681} 34c

\bibitem{strange_max} Braun-Munzinger P, Cleymans J, Oeschler H and
Redlich K 2001 \NP A {\it in print}\\
({\it Preprint} hep-ph/0106066)


\bibitem{ceres1} Agakichiev G {\it et al} CERES collaboration 1995 
\PRL {\bf 75} 1272

\bibitem{ceres2} Agakichiev G {\it et al} CERES collaboration 1998 \PL B {\bf
422} 405
 
\bibitem{rapp} Rapp R and Wambach J 2000 {\it Adv. Nucl. Phys.} {\bf 25} 1
and references therein

\bibitem{schmitz} Schmitz W 2001 Dissertation, University of Heidelberg

\bibitem{damja} Damjanovic S {\it et al} CERES collaboration 2001
{\it Proc. Int. Europhysics Conf., Budapest, Oct. 2001}\\
({\it Preprint} nucl-ex/0111009)

\bibitem{fili} Filimonov K {\it et al}, {\it Proc. INPC 2001 Conference,
Berkeley, August 2001}\\
({\it Preprint} nucl-ex/0109017)

\bibitem{gsi} An International Accelerator facility for Beams of Ions and
Antiprotons, Conceptual Design Report, GSI, Nov. 2001.

\bibitem{specht} Specht H J 2001 {\it Proc. QM2001 conference, Stony Brook,
Jan. 2001} \NP A {\it in print} \\
({\it Preprint} nucl-ex/0111011)

\bibitem{nagle} Nagle J L 2001 {\it Proc. International Nuclear Physics
Conference (INPC 2001), Berkeley, California, July 2001}\\
({\it Preprint} nucl-ex/0109016)

\bibitem{becattini1} Becattini F and Heinz U 1997 \ZP C {\bf 76} 269\\
({\it Preprint} hep-ph/9702274)\\ 
See also: Becattini F {\it Preprint} hep-ph/9701275

\bibitem{redlich} Redlich K 2001 {\it Preprint} hep-ph/0111383 \\ 
Redlich K and Tounsi A {\it Preprint} hep-ph/0111261

\bibitem{E864} Armstrong T A {\it et al} E864 collaboration 2000
\PR C {\bf 61} 064908

\bibitem{E864_1} Armstrong T A {\it et al} E864 collaboration 2000
\PRL {\bf 85} 2685

\bibitem{jphysg} Braun-Munzinger P and Stachel J 1995  \jpg {\bf 21} L17

\bibitem{dover1} 
Baltz A J {\it et al} 1994 \PL B {\bf 325} 7\\ 
Note that eq. (5) of this manuscript contains an error which is 
corrected in our Table\ \ref{probab}.

\end{thebibliography}
\end{document}